\documentclass{an_2c}
\usepackage{psfig}
\begin{document}
\sloppy

\title{Recent ASCA and SAX observations of intermediate BL Lac objects}
\author{J. Siebert\inst{1}, W. Brinkmann\inst{1}, M. Gliozzi\inst{1}, 
S.A. Laurent-Muehleisen\inst{2}\and  M. Matsuoka\inst{3}}
\institute{
MPI f\"ur extraterrestrische Physik, Postfach 1603, 85740 Garching, Germany
\and 
IGPP/LLNL, Livermore, CA 94550, USA
\and 
RIKEN, 2-1 Hirosawa, Wako, 351-01 Saitama, Japan}
\headnote{Astron. Nachr. 320 (1999)}
\maketitle

\section{Introduction}
\vspace*{-0.2cm}

BL Lac objects detected in X-ray and radio surveys show markedly different properties
and are clearly separate in terms of their spectral energy distributions (SEDs), parameterized
by, e.g. $\alpha_{\rm rx}$, the ratio of the 5GHz radio to the 1 keV X-ray flux. 
The unification of BL Lac objects is a long-standing, but still unsolved issue. However, 
recent observational and theoretical progress indicates that all BL Lacs do indeed belong
to one population. New samples derived from the ROSAT All-Sky Survey (RASS) and sensitive radio
catalogs show a continuous distribution in $\alpha_{\rm rx}$ (e.g. Laurent-Muehleisen et al. 
1999). Theoretical work indicates that variations in the peak energy of the synchrotron 
emitting electron population combined with orientation might be able to explain the range
of properties among BL Lac objects (e.g. Celotti 1993; Padovani \& Giommi 1994).
In this contribution we present the broad band X-ray spectra of intermediate BL Lac objects 
(IBLs) discovered in our correlation of the RASS and the 5GHz 87GB radio survey (cf. Brinkmann
et al. 1995, 1997). They show $\alpha_{\rm rx}$ values {\it intermediate} between the previously
known classes of X-ray and radio selected BL Lacs.       
\vspace*{-0.3cm}

\section{Observations and results}
\vspace*{-0.2cm}

\begin{table}
\caption{Results of the spectral analysis.}
\label{results}
\tabcolsep1.ex
\tiny
\begin{tabular}{llllllll} \hline \noalign{\smallskip}
\bf Name & \bf Obs. & $\Gamma_{\bf l}$ & $\Gamma_{\bf h}$ & \bf E$_{\bf br}$ [keV] & \bf F$_{\bf 2-10 keV}$ \\
\noalign{\smallskip} \hline\hline \noalign{\smallskip}
1034$+$5727 & SAX   & $2.10^{+0.15}_{-0.16}$ & $2.42^{+0.26}_{-0.25}$ & 2.0 (fixed) & $0.8\times 10^{-12}$ \\
\noalign{\smallskip}
1055$+$5644 & SAX 1 & $2.34^{+0.20}_{-0.22}$ & $3.21^{+1.25}_{-0.92}$ & 1.75 (fixed) & $0.3\times 10^{-12}$\\
1055$+$5644 & SAX 2 & $2.22^{+0.11}_{-0.13}$ & $2.78^{+0.25}_{-0.25}$ & 1.75 (fixed) & $1.4\times 10^{-12}$\\
1055$+$5644 & ASCA  & $2.23^{+0.11}_{-0.13}$ & $2.76^{+0.09}_{-0.07}$ & $1.75^{+0.21}_{-0.18}$ & $2.9\times 10^{-12}$ \\
\noalign{\smallskip}
1424$+$2401 & ASCA  & $2.78^{+0.04}_{-0.04}$ & $1.22^{+0.90}_{-0.73}$ & 5.0 (fixed) & $1.9\times 10^{-12}$\\
\noalign{\smallskip}
1741$+$1936 & SAX   & $1.71^{+0.20}_{-0.15}$ & $2.10^{+0.08}_{-0.07}$ & $1.90^{+0.67}_{-0.69}$ & $5.4\times 10^{-12}$\\
\noalign{\smallskip} \hline
\end{tabular}
\end{table}

In total we performed 6 observations of 4 IBLs with both ASCA and SAX. Table~\ref{results} 
summarizes the results from our spectral analysis. In general, the hard X-ray spectra of IBLs
are very steep with photon indices ranging from 2.1 to 2.8. Furthermore, a broken power law 
provides a significantly better description of the X-ray spectra of IBLs than a simple power 
law. The X-ray spectra get flatter towards lower energies. In addition to that, the spectrum of 
1424+2401 also seems to flatten above 5 keV. This may indicate that a flat inverse Compton 
component starts to dominate the X-ray spectrum. The X-ray flux from 1055+5644 varied by 
almost an order of magnitude between our three observations and the ASCA data show a 50\% decrease 
within one day. Interestingly, no significant spectral changes are found, which might provide
important constraints on electron injection and cooling processes.   

\begin{figure}
\hspace*{0.4cm}
\psfig{file=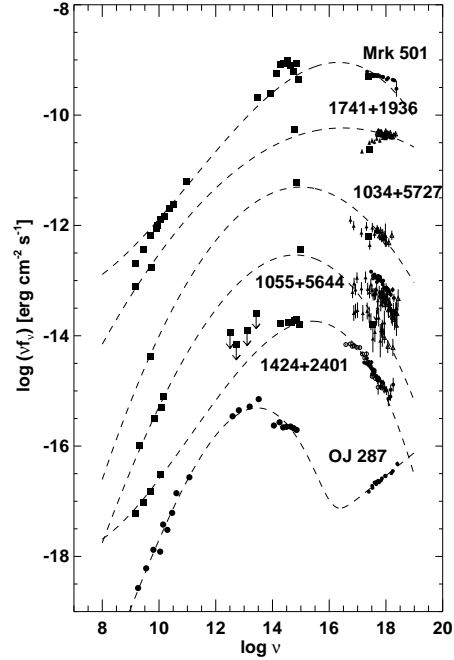,width=6.5cm}
\caption{The spectral energy distribution of various IBLs. The normalization is arbitrary.}
\label{sed}
\end{figure}

In Fig.~\ref{sed} we present the SEDs of our IBLs and compare them to Mrk 501 and
OJ 287, typical X-ray and radio selected BL lac objects, respectively. 
The steep and convex X-ray spectra as well as the SEDs of IBLs both confirm their 
intermediate nature. The SEDs are characterized by a synchrotron peak in the UV/soft 
X-ray band. In hard X-rays we obviously observe the steep tail of the synchrotron component. 
Variability and uncertainties in the absorption correction might explain the discontinuities
between the X-ray and the optical band, in particular for 1741+1936.
In one case (1424+2401) we might see the onset of the flat inverse-Compton component.

\end{document}